\documentclass[9pt,twocolumn,twoside]{pnas-new}
\usepackage{arydshln} %for the dashed div.lines in table

\templatetype{pnasresearcharticle} % Choose template 
\title{Improved Lemaitre-Tolman model and the mass and turn-around radius in group of galaxies}

\author[a,b]{Antonino Del Popolo}
\author[c,2]{Maksym Deliyergiyev} 
\author[d]{Man Ho Chan}

\affil[a]{Dipartimento di Fisica e Astronomia, University Of Catania, Viale Andrea Doria 6, 95125, Catania, Italy}
\affil[b]{Institute of Astronomy, Russian Academy of Sciences, Pyatnitskaya str. 48, 119017 Moscow, Russia}
\affil[c]{Département de Physique Nucléaire et Corpusculaire, University of Geneva, CH-1211 Geneve 4, Switzerland}
\affil[d]{Department of Science and Environmental Studies, The Education University of Hong Kong, Tai Po, New Territories, Hong Kong}

\leadauthor{A. Del Popolo} 

\correspondingauthor{\textsuperscript{2} E-mail: maksym.deliyergiyev@unige.ch}


\begin{abstract}
We extended the modified Lemaitre-Tolman model Ref.\citep{Peirani2006,Peirani2008} taking into account the effect of angular momentum
and dynamical friction. The inclusion of these quantities in the equation of motion modifies the evolution of a perturbation, initially moving with the Hubble flow. Solving the equation of motions we got the relationships between mass, $M$, and the turn-around radius, $R_0$. Knowing $R_0$, the quoted relation allows the determination of the mass of the object studied. The relationships for the case in which also the angular momentum is taken into account gives a mass $\simeq 90$ \% larger than the standard Lemaitre-Tolman model, and two times the value of the standard Lemaitre-Tolman model, in the case also dynamical friction is taken into account. As a second step, we found relationships between the velocity, $v$, and radius, $R$, and fitted them to data of the Local Group, M81, NGC 253, IC342, CenA/M83, and to the Virgo clusters obtained by Ref.\citep{Peirani2006,Peirani2008}. This allowed us to find optimized values of the mass and Hubble constant of the objects studied. The fit gives values of the masses smaller with respect to the $M-R_0$ relationship method, but in any case  30-40\%  larger than the $v-R$ relationship obtained from the standard Lemaitre-Tolman model. Differently from mass, the Hubble parameter becomes smaller with respect to the standard Lemaitre-Tolman model, when angular momentum, and dynamical friction are introduced. This is in agreement with Ref.\citep{Peirani2006,Peirani2008}, who improved the standard Lemaitre-Tolman model taking into account the cosmological constant.

Finally, we used the mass, $M$, and $R_0$ of the studied objects to put constraints to the dark energy equation of state parameter, $w$. Comparison with previous studies show different constraints on $w$. 
\end{abstract}

\begin{document}

\maketitle
\thispagestyle{firststyle}
\ifthenelse{\boolean{shortarticle}}{\ifthenelse{\boolean{singlecolumn}}{\abscontentformatted}{\abscontent}}{}

% If your first paragraph (i.e. with the \dropcap) contains a list environment (quote, quotation, theorem, definition, enumerate, itemize...), the line after the list may have some extra indentation. If this is the case, add \parshape=0 to the end of the list environment.
%\dropcap{T}his PNAS journal template is provided to help you write your work in the correct journal format.  Instructions for use are provided below. 

%Note: please start your introduction without including the word ``Introduction'' as a section heading (except for math articles in the Physical Sciences section); this heading is implied in the first paragraphs. 

%%%%%%%%%%%%%%%%%%%%%%%%%%%%    
%%%%%%%%%%%%%%%%%%%%%%%%%%%% 
%\section*{Introduction}

\dropcap{W}hile the mass-to-light ($M/L$) ratios of group of galaxies was in the past estimated through the virial theorem to be typically of the order of $\simeq 170 M_{\odot}/L_{B,M_{\odot}}$ Ref.\cite{Huchra1982}, new measurements based on high quality data, and estimating methods different from the Virial theorem Ref.\cite{Karachentsev2004} give much smaller results in the range $10-30 M_{\odot}/L_{B,M_{\odot}}$. This means that the local matter density should be a fraction of the global one. It is well known that the virial theorem gives reliable results if the system is in dynamical equilibrium. This condition is often assumed if the crossing time is less than the Hubble time. This assumption has been shown to be often not correct by Ref.\cite{Niemi2007}, whose analysis showed that there is no correlation between the virial ratio $\frac{2T}{W}$, being $T$, and $W$ the kinetic and potential energy, and the crossing time. By means of methods used by observers, Ref.\cite{Niemi2007} showed that $\simeq 20 \%$ of the studied groups were not gravitationally bound.   
Ref.\cite{LyndenBell1981} and Ref.\cite{Sandage1986} proposed an alternative approach to the virial theorem based on the Lemaitre-Tolman (LT) model Ref.\cite{Lemaitre1933,Tolman1934} giving a good description of a central core gravitationally bound located inside an homogeneous region whose density decreases till reaching the background value. The model describes the evolution of the system in a similar way to that done by the spherical collapse model. Considering a shell of given radius containing a mass $M$, it initially expands following the Hubble flow. When the density overcomes a critical value the shell reaches a maximum radius, known as turn-around radius, $R_0$, characterized by zero velocity, and collapses. Then in the LT model there is a central region in equilibrium, surrounded by a region which reaches its maximum expansion and collapses, and a zero totally energy region constituted by shells still bound to the structure and unbound ones. Because of its characteristics, the LT model gives a good description of a group of galaxies dominated by one or two central galaxies embedded into a cloud of smaller ones. If using the velocity field around the main bodies allows the determination of the turn-around radius $R_0$, the mass can be obtained through the relation 
\begin{equation}
\label{eq:LT}
M=\frac{\pi^2 R_0^3}{8GT_0^2}
\end{equation}
Refs.\cite{Sandage1986,Peirani2006,Peirani2008}, where $T_0$ is the age of the universe. The quoted model was applied to the local group Ref.\cite{Sandage1986} and to the Virgo cluster Refs.\cite{Hoffman1980, Tully1984,Teerikorpi1992}. The model was modified taking into account the cosmological constant by Ref.\cite{Peirani2006,Peirani2008} applying it to the Virgo cluster, the pair M31-MW, M81, the Centaurus A-M83 group, the IC342/Maffei-I group, and the NGC 253 group. 
As shown in Refs.\cite{Peirani2006,Peirani2008} the introduction of the cosmological constant modifies the mass, $M$, turn-around radius, $R_0$, relation. As a consequence for a given $R_0$, the value of the mass of the system is $\simeq 30\%$ larger with respect to \eqref{eq:LT} Refs.\citep{Peirani2006,Peirani2008}, while the Hubble constant of the modified model is smaller than the standard LT.

In order to obtain the mass of the previously quoted objects, Refs.\cite{Peirani2006,Peirani2008}, differently from Ref.\cite{Sandage1986}, did not use the standard LT (SLT) $M-R_0$ relation (\eqref{eq:LT}). 
They built up a velocity-distance relationship, $v-R$, describing the kinematic status of the systems studied. Knowing the values of $v$, and $r$ for the members of the groups studied, the mass of the group, $M$, and the Hubble parameter can be obtained by means of a non-linear fit of the $v-R$ relation to the data. 

In the present paper, we will further extend the modified Lemaitre-Tolman (MLT model)
by taking into account the effect of angular momentum (JLT model) and dynamical friction (J$\eta$LT model). The effect of these two quantities on the spherical collapse model (SCM) and its effect on the clusters of galaxies structure and evolution, the turn-around, the threshold of collapse, their mass function, their mass-temperature relation, have been studied in Refs.\cite{DelPopolo1998,DelPopolo1999,DelPopolo2000,DelPopolo2006,DelPopolo2006a,DelPopolo2006b,DelPopolo2017,DelPopolo2019,DelPopolo2020}. 

Similarly to Ref.\cite{Peirani2006,Peirani2008}, we will find the $v-R$ relation by solving the equation of the SCM, and then fit it to the data of the Virgo cluster, the pair M31-MW, M81, the Centaurus A-M83 group, the IC342/Maffei-I group, and the NGC 253 group.

The paper is organized as follows. In Section~\ref{sec:Model}, we introduce the model, and solve it. In Section~\ref{sec:VR_relation}, we find the velocity-radius relation for the JLT, and J$\eta$LT models. 
In Section~\ref{sec:ApplicationNearGroups}, 
we applied the $v-R$ relation to groups and clusters of galaxies.
In Section~\ref{sec:EffectCosmologicalConst}, we studied the impact of the angular momentum and dynamical friction on the $M-R_0$ relation.  In Section~\ref{sec:DMconstraints}, we showed how the obtained values of $M$ and $R_0$ may constrain the dark energy equation of state parameter, $w$.
Section~\ref{sec:Conclusions} is devoted to conclusions.

%%%%%%%%%%%%%%%%%%%%%%%%%%%%%%%%%%%%%%%%%%%%%%%%
%%%%%%%%%%%%%%%%%%%%%%%%%%%%%%%%%%%%%%%%%%%%%%%%
\section*{Model}
\label{sec:Model}

The simplest form of the SCM was introduced by Ref.\cite{Gunn1972}. It is a simple and popular method to study analytically the non-linear evolution of perturbations of dark matter (DM) and dark energy (DE). As previously reported, the model describes the evolution of a spherical symmetric over density which initially expands with the Hubble flow, then detaches from it, when the density overcomes a critical value, reaches a maximum radius, dubbed turn-around radius, and finally collapse and virialize. SCM is a very simple model assuming that matter moves in a radial fashion Ref.\cite{Gunn1972, Silk1974ApJ.193.525S, Gunn1977}. 
Tidal angular momentum Ref.\cite{Peebles1969,White1984}, random angular momentum Refs.\cite{Ryden1987,Ryden1988,Williams2004}, dynamical friction (Refs.\cite{AntonuccioDelogu1994,Delpopolo2009}), etc., are not taken into account. Later the SCM was improved in several papers Ref.\cite{Fillmore1984,Bertschinger1985,Hoffman1985,Ryden1987,Subramanian2000,Ascasibar2004,Williams2004},  
adding the cosmological constant Ref.\cite{Lahav1991}, and tidal and random angular momentum\cite{Ryden1987,Gurevich1988a,Gurevich1988b,White1992,Sikivie1997,Nusser2001,Hiotelis2002,  
LeDelliou2003,Ascasibar2004,Williams2004,Zukin2010}\footnote{Particles angular momenta is randomly distributed in random such that the mean angular momentum at any point in space is zero Ref.\cite{White1992,Nusser2001} then conserving spherical symmetry and angular momentum.}. Dynamical friction was studied in Refs.\cite{AntonuccioDelogu1994,Delpopolo2009}, while Refs.\cite{Hoffman1986,Hoffman1989,Zaroubi1993} discussed the role of shear in the gravitational collapse.

The SCM with negligible DE perturbations was extensively investigated in literature 
Refs.\cite[see, e.g.][]{Bernardeau1994,Bardeen1986,Ohta2003,Ohta2004,Basilakos2009,Pace2010,Basilakos2010}, 
while DE fluid perturbation were taken into account in Refs.\cite[see][]{Mota2004,Nunes2006,Abramo2007,Abramo2008,Abramo2009a,Abramo2009b,Creminelli2010,Basse2011,Batista2013}. 
Using the non-linear differential equations for the evolution of the matter density contrast derived from Newtonian hydrodynamics in Ref.\cite{Pace2010}, Ref.\cite{DelPopolo2013b} showed that the parameters of the SCM become mass dependent. 

Refs.\cite{DelPopolo2013a,DelPopolo2013b} studied the effects of shear and rotation in smooth DE models.
The effects of shear and rotation were investigated in Refs.\cite{DelPopolo2013a,DelPopolo2013b} for smooth DE models, Ref.\cite{Pace2014b} in clustering DE cosmologies, and 
Ref.\cite{DelPopolo2013c} in Chaplygin cosmologies.

In this paper, we are interested in describing a system constituted by a dominant mass concentration, and satellites that are not  contributing significantly to the group mass, and that further mass accretion is neglected.  

The equation of motion of the system may be obtained as follows. We consider some gravitationally growing mass concentration collecting into a potential well. 
Let us assume that the probability of a particle, located at $[r,r+{\rm d}r]$, 
having angular momentum  $L=r v_{\theta}$, 
defined in the range $[L,L+{\rm d}L]$, with velocity $v_r={\dot r}$, 
defined in the range $[v_r,v_r+{\rm d}v_r]$, has the following form
\begin{equation}
{\rm d}P=f(L,r,v_r,t){\rm d}L{\rm d}v_r{\rm d}r.
\end{equation}
The term $L$ takes into account ordered angular momentum generated by tidal torques and random angular momentum (see Appendix C.2 of Ref.\cite{Delpopolo2009}). The radial acceleration of the particle Refs.\cite{Peebles1993,Bartlett1993,Lahav1991,DelPopolo1998,DelPopolo1999} is: 
\begin{equation}
\label{eq:coll}
 \frac{{\rm d}v_R}{{\rm d}t} = -\frac{GM}{R^2} + \frac{L^2(R)}{M^{2}R^3} + \frac{\Lambda}{3}R -
 \eta\frac{{\rm d}R}{{\rm d}t}\,,
\end{equation}
with $\Lambda$ being the cosmological constant and $\eta$ the dynamical friction coefficient. The previous equation can be obtained via Liouville's theorem Ref.\cite{DelPopolo1999}. The last term, the dynamical friction force per unit mass, $\eta$, is explicitly given in Ref.\cite{Delpopolo2009} (Appendix D, Eq. D5). A similar equation (excluding the dynamical friction term) was obtained by several authors Refs.\cite[e.g.,][]{Fosalba1998a,Engineer2000,DelPopolo2013b}) and generalized to smooth DE models in Ref.\cite{Pace2019}.
%====================================================
\begin{figure}[t]
 \centering
\includegraphics[scale=0.45]{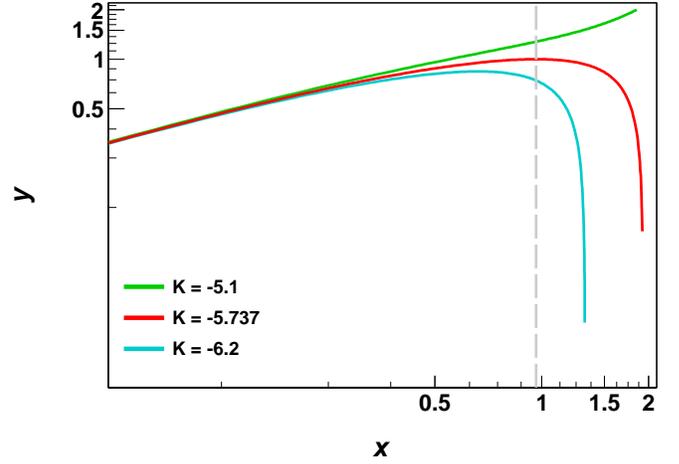}
 \caption{
 Evolution of shell radius for different values of $K$. The red, cyan, and green lines correspond to $K=-5.737$, $K=-6.2$, and $K=-5.1$, respectively.
}
\label{fig:RadiusEvoluion_diff_K}
\end{figure}
%====================================================

In terms of the specific angular momentum $J=\frac{L}{M}$, and $\Omega_{\Lambda}=\frac{\rho_{\Lambda}}{\rho_c}$, where $\rho_c$ is the critical density, \eqref{eq:coll} can be written as
\begin{equation}
\label{eq:coll1}
 \frac{{\rm d}v_R}{{\rm d}t} = -\frac{GM}{R^2} + \frac{J^2}{R^3}-\frac{1+3w}{2} \Omega_{\Lambda} H_0^2 \left(\frac{a_0}{a}\right)^{3(1+w)} R
-\eta\frac{{\rm d}R}{{\rm d}t}\,,
\end{equation}
where $w$ is the DE equation of state (EoS) parameter. DE is modeled by a fluid with an EoS $P=w \rho$, where $\rho$ is the energy density. $a$ is the expansion parameter. \eqref{eq:coll1} satisfies equation
\begin{equation}
\label{eq:hubb}
H=\frac{\dot a}{a}=H_0 \sqrt{\Omega_{\rm m} \left(\frac{a_0}{a}\right)^3+\Omega_{\Lambda}\left(\frac{a_0}{a}\right)^{3(1+w)} }.
\end{equation}

In the following, we will treat the case $w=-1$, in other words we assume that DE is the cosmological constant. With this assumption, and assuming that $J=k R^{\alpha}$, with $\alpha=1$, in agreement with Ref.\cite{Bullock2001}\footnote{In that paper $\alpha=1.1 \pm 0.3$}, and $k$ constant. In terms of the variables $y=R/R_0$, $t=x/H_0$, \eqref{eq:coll1}, and \eqref{eq:hubb} can be written as
\begin{equation}
\label{eq:princ}
\frac{d^2y}{dx^2}=-\frac{A}{2y^2}+\Omega_{\Lambda} y+\frac{K_j}{y}-\frac{\eta}{H_0}\frac{dy}{dx},
\end{equation}
where $K_j=k\frac{1}{(H_0R_0)^2}$, $A=\frac{2GM}{H_0^2 R_0^3}$, and
\begin{equation}
\label{eq:hubb1}
H=H_0 \sqrt{\Omega_m \left(\frac{a_0}{a}\right)^3 +\Omega_{\Lambda}}.
\end{equation}
\eqref{eq:princ} has a first integral, given by
\begin{align} 
\label{eq:princ1}
u^2&=\left(\frac{dy}{dx}\right)^2\nonumber\\
&=\frac{A}{y}+\Omega_{\Lambda} y^2+2K_j \log{y}-2\frac{\eta}{H_0} \int \left(\frac{dy}{dx}\right)^2 dx+K
\end{align}
where $K=\frac{2E}{(H_0R_0)^2}$, and $E$ is the energy per unit mass of a shell.

Eqs.(\ref{eq:hubb}), and (\ref{eq:princ}) where solved as described in Ref.\cite{Peirani2006, Peirani2008}. There are a couple of ways of doing that. A first way, is to obtain the value of the scale parameter and the corresponding time for a given redshift. At high redshift, the gravitational term dominates and through a Taylor expansion one can get the initial conditions. In order to get the parameter $A$, it is varied until the condition $\frac{dy}{dx}=0$, and $y=1$ are satisfied. A second way to get $A$, is to use the equation for the velocity (\eqref{eq:princ1}).

%=================================================
\begin{table}[t]
	\begin{center}
		\begin{tabular}{l|ccccc}
			\hline
			model
			&$\eta/H_0$
			&\multicolumn{1}{c}{$K_{J}$}
			&\multicolumn{1}{c}{$b$}	
			&\multicolumn{1}{c}{$n$}
			&\multicolumn{1}{c}{$A$}\\	
			\hline
			\multicolumn{1}{c}{MLT} 
			&\multicolumn{1}{|c}{--} 
			&\multicolumn{1}{l}{$0.0$}  
			&\multicolumn{1}{l}{$1.4054$}
			&\multicolumn{1}{l}{$0.6293$} 
			&\multicolumn{1}{l}{$3.6575$}\\						
			\multicolumn{1}{c}{JLT} 
			&\multicolumn{1}{|c}{--} 
			&\multicolumn{1}{l}{$0.78$} 
			&\multicolumn{1}{l}{$1.3759$}
			&\multicolumn{1}{l}{$0.7549$} 
			&\multicolumn{1}{l}{$5.0370$}\\	
			\multicolumn{1}{c}{J$\eta$LT } 
			&\multicolumn{1}{|c}{$0.5$} 
			&\multicolumn{1}{l}{$0.78$} 
			&\multicolumn{1}{l}{$1.3436$}
			&\multicolumn{1}{l}{$0.9107$} 
			&\multicolumn{1}{l}{$6.0500$}\\																
			\hline								
		\end{tabular}
	\end{center}
	\caption{
	The constant $A$, and the fitting parameters $b$, and $n$ of the velocity-distance ($v-R$) relations,
	for the MLT, the JLT, and the J$\eta$LT model.
}
\label{tab:fit_param}	
\end{table} 
%=================================================

Let's show this second method in the case cosmological constant, and angular momentum are present (JLT case)
\begin{equation} 
\label{eq:princ2}
\frac{d^2y}{dx^2}=-\frac{A}{2y^2}+\Omega_{\Lambda} y+\frac{K_j}{y}
\end{equation}
having the first integral
\begin{align} 
\label{eq:princ3}
u^2=\left(\frac{dy}{dx}\right)^2 =\frac{A}{y}+\Omega_{\Lambda} y^2+2K_j \log{y}+K
\end{align}
At the turn-around point \eqref{eq:princ3} gives: $K=-A-\Omega_{\Lambda}$.

At high redshifts ($z=1000$), or $y\ll 1$, as was described the gravitational term dominates, and by a Taylor expansion one gets the  
relation $y \simeq (\frac{9A}{4})^{1/3} x^{2/3}$. Assuming an initial value of $y$, $y_i=0.001$, corresponding approximately to 1 kpc, the initial time $x_i$ can be obtained. The initial value of the velocity $u_i$ can be obtained, when $A$ is known, through \eqref{eq:princ3}, recalling that $y_i=0.001$. The value of $A$ is obtained as follows. \eqref{eq:princ3} can be written as
\begin{equation} 
\label{eq:x}
x=\int_{y_i}^1 \frac{dy}{
\sqrt{\frac{A}{y}+\Omega_{\Lambda} y^2+\frac{K_j}{y}-A-\Omega_{\Lambda}}
}
\end{equation}
\eqref{eq:hubb1}, recalling that $\frac{a_0}{a}=1+z$, can be written as
\begin{equation} 
\label{eq:y}
x(y=1)=\int_0^{\infty} \frac{dz}{
(1+z)\sqrt{\Omega_{\Lambda}+\Omega_{\rm m} (1+z)^3}
}=0.964.
\end{equation}
For $\Omega_{\Lambda}=0.7$, $\Omega_{\rm m}=0.3$, $K_j=0.78$\footnote{The value of $K_j$ was obtained recalling that term related to angular momentum, L, in \eqref{eq:coll1}, is given by $\frac{L^2}{M^2 R^3}$.}, $x=0.964$, \eqref{eq:x} can be solved to get $A=5.037$. In the case, $K_j=0$, $A=3.6575$, and if $K_j=0$, $\Omega_{\Lambda}=0$, the SLT gives $A=2.655$. 

In the case J$\eta$LT (Eqs.(\ref{eq:princ})-(\ref{eq:princ1})), $A$ can be obtained similarly to the previous case (JLT)
solving numerically \eqref{eq:princ1} with the initial condition on $y_i$, and varying $A$ until the condition 
$\frac{dy}{dx}=0$, and $y=1$ are satisfied. Similarly, we can solve \eqref{eq:princ} with the initial condition $y_i$, and varying $A$ until the condition $\frac{dy}{dx}=0$, and $y=1$ are satisfied.
In this way, one gets $A=6.05$.

Now, we show the solution for the case JLT. \eqref{eq:princ2} can be solved with the conditions $y_i=0.001$, and 
\begin{equation}
\begin{split}
\label{eq:u0}
u(0)&=\sqrt{\frac{5.037}{y_i}+0.7 y_i^2+2 K_J \log{y_i}-0.7-5.037}\\
&=70.833943.
\end{split}
\end{equation}
In Fig.\ref{fig:RadiusEvoluion_diff_K}, we plot the result of the solution. The red line corresponds to the case $K=-A-\Omega_{\Lambda}=-5.737$, being $A=5.037$. 
This solution is the one that has just reached the maximum expansion, or turn-around, and the collapse happens in $\simeq 13.8$ Gyr. 
The cyan line is characterized by $K=-6.2$. It reached the turn-around in the past. 
Turn-around will happen only for $K<-5.56812$, for larger values the collapse will never occur, as the case of the green line characterized by $K=-5.1$.

%====================================================
\begin{figure*}[!ht]
 \centering
\includegraphics[scale=0.8]{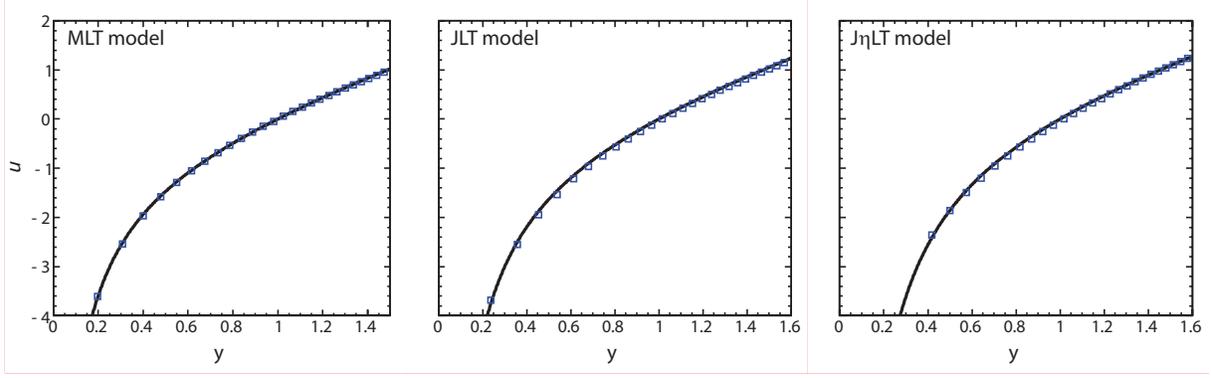}
 \caption{
 The velocity profile in the three cases studied. The left panel shows the MLT model. The central panel, the JLT. The right panel, the J$\eta$LT case. Data points are obtained as follows. Fig. 1, shows some solutions of Eq. 6 for different values of K: K = -6.2, -5.737, and -5.1. We obtained several solution relative to different values of K: $K_1$, $K_2$, $K_3$..... The intersection of each solution with the vertical axis $x = 0.964$, gives a corresponding value of y: $y_1$, $y_2$, $y_3$,.... The solution of Eq. 6 for different values of K at $x = 0.964$ gives also $u_1$, $u_2$, $u_3$....  All this gives us a pair of values $(y_{i},u_{i})$ for each intersection of the vertical line with the curves. The solid black lines are the fit to the points in the examined models. 
}
\label{fig:agnularMom_dynFric}
\end{figure*}
%====================================================

%%%%%%%%%%%%%%%%%%%%%%%%%%%%%%%%%%%%%%%%%%%%%%%%
%%%%%%%%%%%%%%%%%%%%%%%%%%%%%%%%%%%%%%%%%%%%%%%%
\section*{The velocity-radius relation}
\label{sec:VR_relation}

In order to get the mass, and turn-around radius of some groups of galaxies, we will find a relation between the velocity, and radius, $v-R$, that will be fitted to the data. The $v-R$ relation is obtained as follows. Let's consider Fig.\ref{fig:RadiusEvoluion_diff_K}. The vertical line corresponds to $x=0.964$. Its intersection with the curves, solution of the equations described in the previous section, gives the value $y(x)=y(0.964)$. The solution of the equations of the previous section, also gives the velocity, allowing us to find $u(x)=u(0.964)$. We will get a couple of value $(y,u)$ for each intersection of the vertical line with the curves (see Fig.\ref{fig:agnularMom_dynFric} caption for an extended description). 
This allows us to find a series of points that can be fitted with a relation of the form $u=-b/y^n+by$. For example in the case of the MLT, we get
\begin{equation}
v=-\frac{1.4054}{y^n}+1.4054y
\end{equation}
where $n=0.6293$. 
This can be written in terms of the physical units as
\begin{equation} 
\label{eq:recov}
v(R)=-b H_0 R_0 \left(\frac{R_0}{R}\right)^n+b H_0 R
\end{equation}
where $b=1.4054$. Substituting in this equation, $R_0=(\frac{2GM}{H_0^2})^\frac{1}{3}$, we get
\begin{equation}
v(R)=-b\frac{H_0}{R^n} \left(\frac{2GM}{A H_0^2}\right)^\frac{n+1}{3}+bH_0R
\end{equation}
or
\begin{equation}
\label{eq:Om}
v(R)=-\frac{-1.013 H_0}{R^n} \left(\frac{GM}{H_0^2}\right)^\frac{n+1}{3}+1.4054 H_0 R
\end{equation}
This relation is slightly different from that obtained by Ref.\cite{Peirani2008}, probably due to the noteworthy sensitivity of the solution of the equation to initial conditions, and to the fact we used more digits in the initial condition for $u(0)$
\footnote{This was confirmed via a private discussion with one of the authors of Ref.\cite{Peirani2008}, namely de Freitas Pacheco.}. 
 
For this reason, in the rest of the paper, we also consider the MLT case, already studied by Ref.\cite{Peirani2008}.
In a similar way, we can obtain the $v-R$ relation in the case the of the JLT model
\begin{equation} 
\label{eq:j}
v(R)=-\frac{-0.80155 H_0}{R^n} \left(\frac{GM}{H_0^2}\right)^\frac{n+1}{3}+1.3759 H_0 R
\end{equation}
where $n=0.7549$, and in the complete case (cosmological constant, angular momentum, and dynamical friction)
\begin{equation}
\label{eq:eta}
v(R)=-\frac{-0.66385 H_0}{R^n} \left(\frac{GM}{H_0^2}\right)^\frac{n+1}{3}+1.3436 H_0 R
\end{equation}
where $n=0.9107$. In Fig.\ref{fig:agnularMom_dynFric} we plot, from left to right, the velocity profile of the MLT, JLT, and J$\eta$LT cases, using adimensional variables.

All the previous equations satisfy the condition $v(R_0)=0$. In the following, we will apply \eqref{eq:eta}, related to the J$\eta$LT model to some groups of galaxies and clusters. In Table \ref{tab:fit_param}, we summarize the parameters of the different models that were described in this paper. The first line corresponds to the MLT model. The second line to the JLT model, and, the last line to the J$\eta$LT case,

Table \ref{tab:fit_param}, as well as Fig.\ref{fig:agnularMom_dynFric} shows that including the angular momentum, and dynamical friction steepens the velocity profile, and increases the parameter $A$. This means that for a given $R_0$ the mass of the structure increases, while the radius of the zero-gravity surface decreases.

%%%%%%%%%%%%%%%%%%%%%%%%%%%%%%%%%%%%%%%%%%%%%%%%
%%%%%%%%%%%%%%%%%%%%%%%%%%%%%%%%%%%%%%%%%%%%%%%%
\section*{Application to near groups and clusters of galaxies}
\label{sec:ApplicationNearGroups}

Now, we will apply Eqs.(\ref{eq:Om}-\ref{eq:eta}) to near groups and a cluster of galaxies. To this aim, we need for each galaxy its velocity and distance with respect to center of mass. We will use data obtained by Refs.\cite{Peirani2006, Peirani2008}. Velocities were transformed from the heliocentric to the Local Group rest frame. The distance can be written as
\begin{equation}
R=\sqrt{D^2+D_g^2-2D D_g \cos{\theta}}
\end{equation}
where the angle $\theta$ is the angle between the center of mass and the galaxy, $D$ the distance from the galaxy to the center of mass, and $D_g$ is the distance to the galaxy. 
Indicating with $V$, and $V_g$ the center of mass velocities, and that of the galaxy with respect the Local Group rest frame, {the velocity difference along the radial direction between both object is}
\begin{equation}
V(R)=V_g \cos{\alpha}-V\cos{\beta}
\end{equation}
being $\alpha=\frac{D \sin_{\theta}}{D_g-D \cos{\theta}}$, and $\beta=\alpha+\theta$.

Since in the list given by Ref.\cite{Peirani2008} unbound objects, and uncertain distances and velocities were excluded, an error of 10\% was considered for velocities and distances by Ref.\cite{Peirani2008}. This value of uncertainty is a weighted mean of data including measurement errors and data reported without errors
 %Ref.\cite{Karachentsev2002v,Karachentsev2002a,Karachentsev2007}
 Ref.\cite{Karachentsev2002a,Karachentsev2007}.

In the case of the group M31-MW, the data were obtained by Ref.\cite{Peirani2006} from Ref.\cite{Karachentsev2002} data. We used the data of Ref.\cite{Peirani2006} also for the case of the Virgo cluster. 

\begin{figure*}
  \centering
   \includegraphics[scale=0.43]{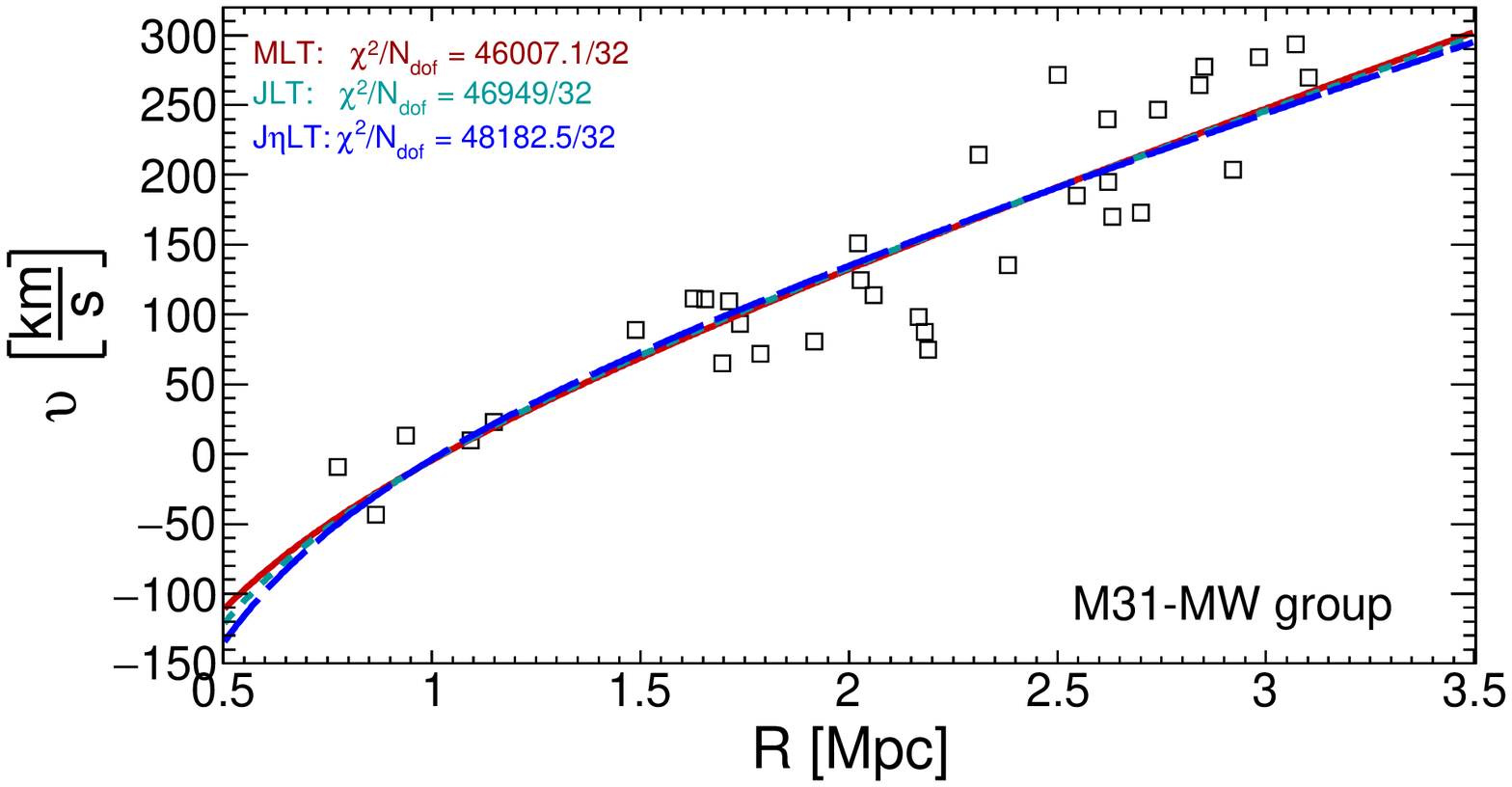} 
   \includegraphics[scale=0.43]{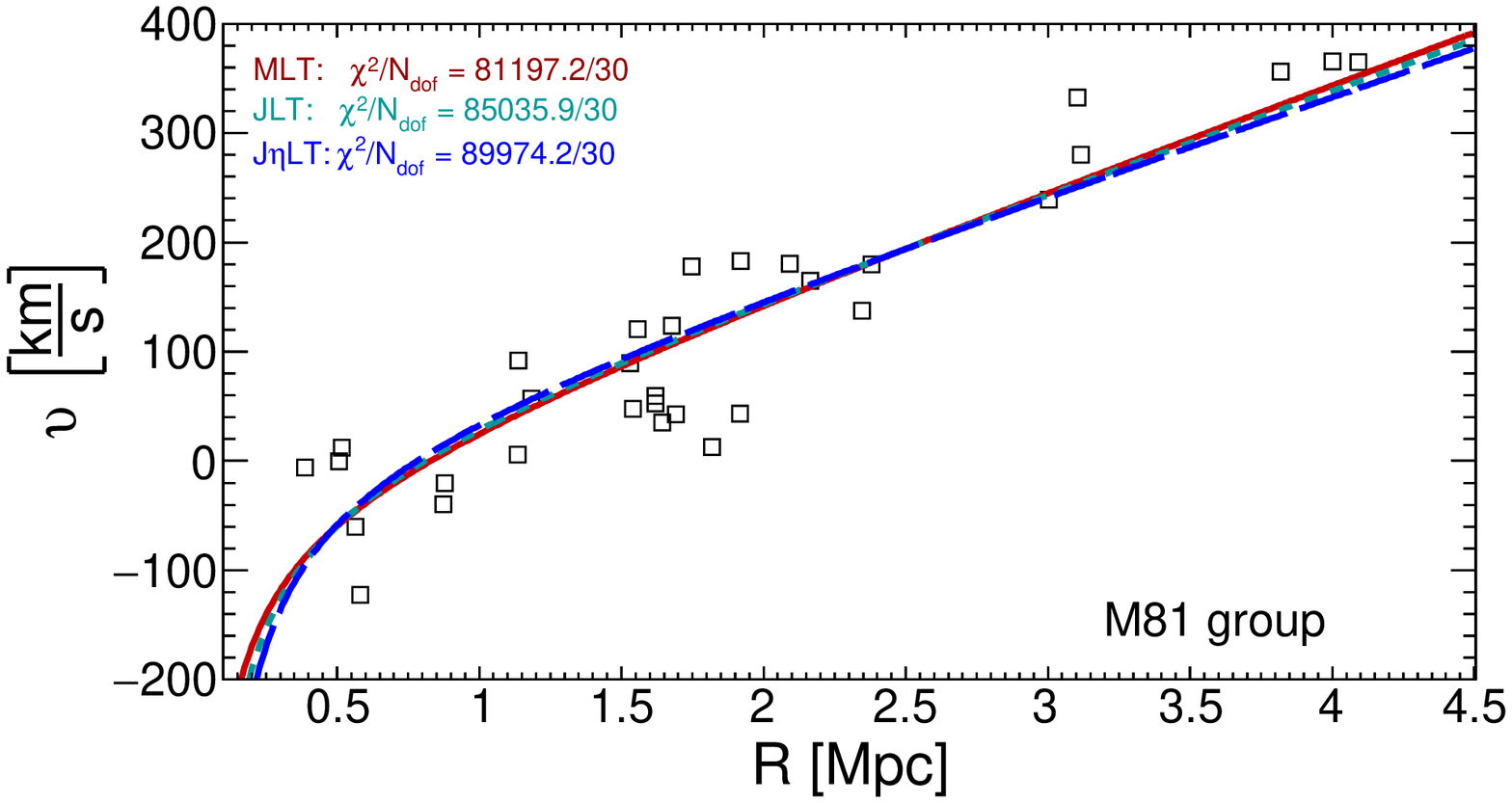} 

   \includegraphics[scale=0.43]{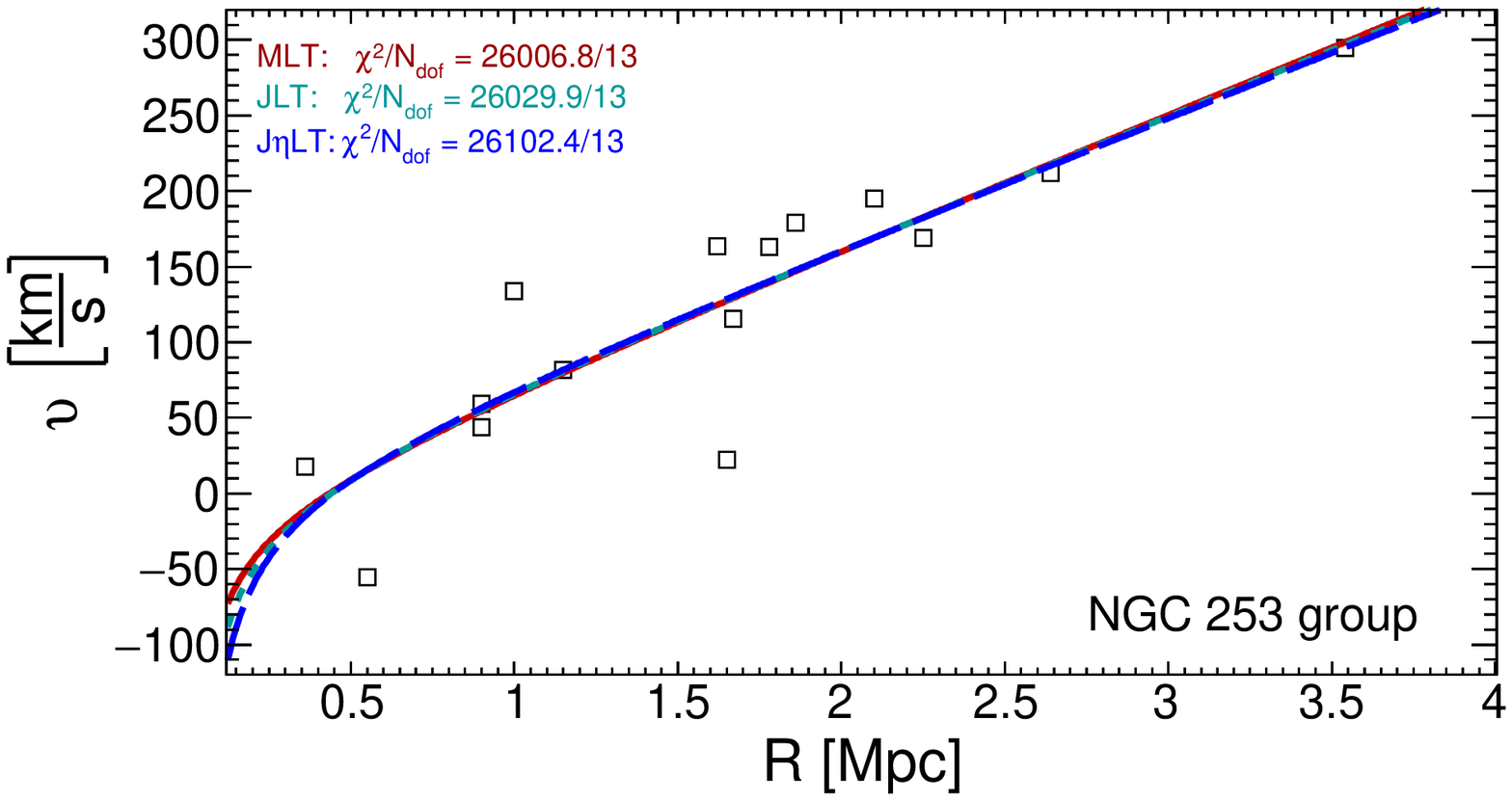} 
   \includegraphics[scale=0.43]{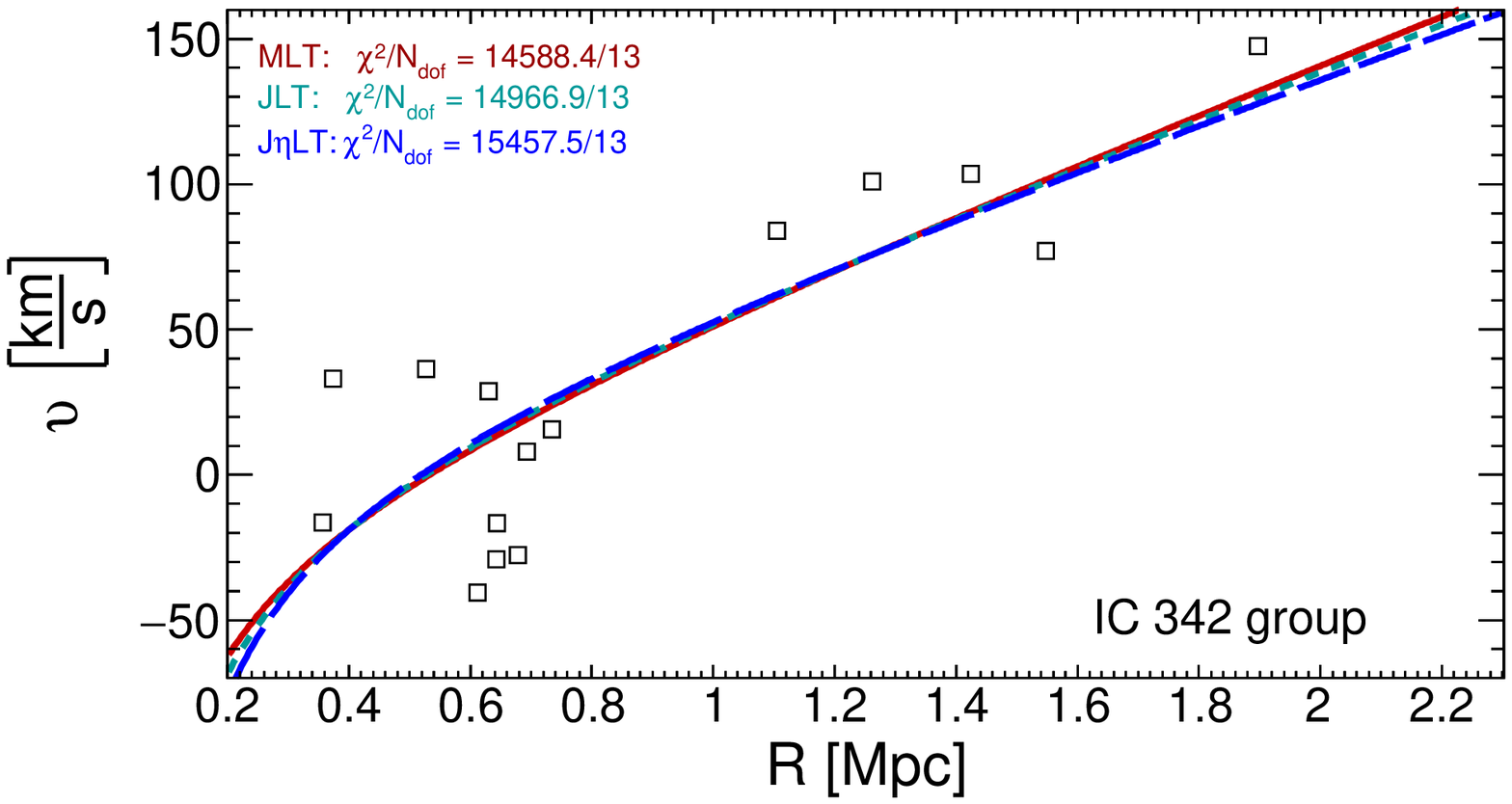} 

   \includegraphics[scale=0.43]{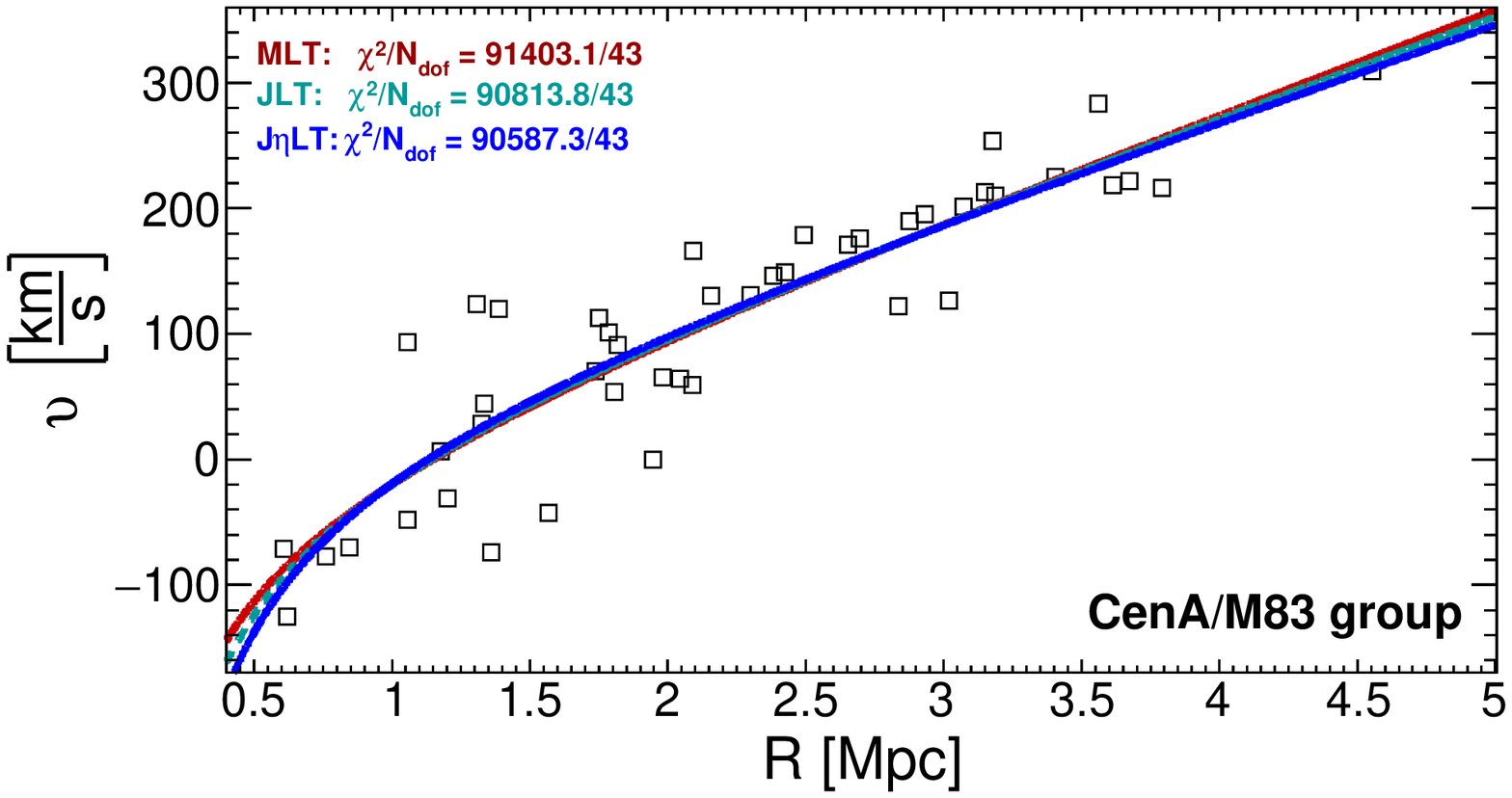} 
   \includegraphics[scale=0.43]{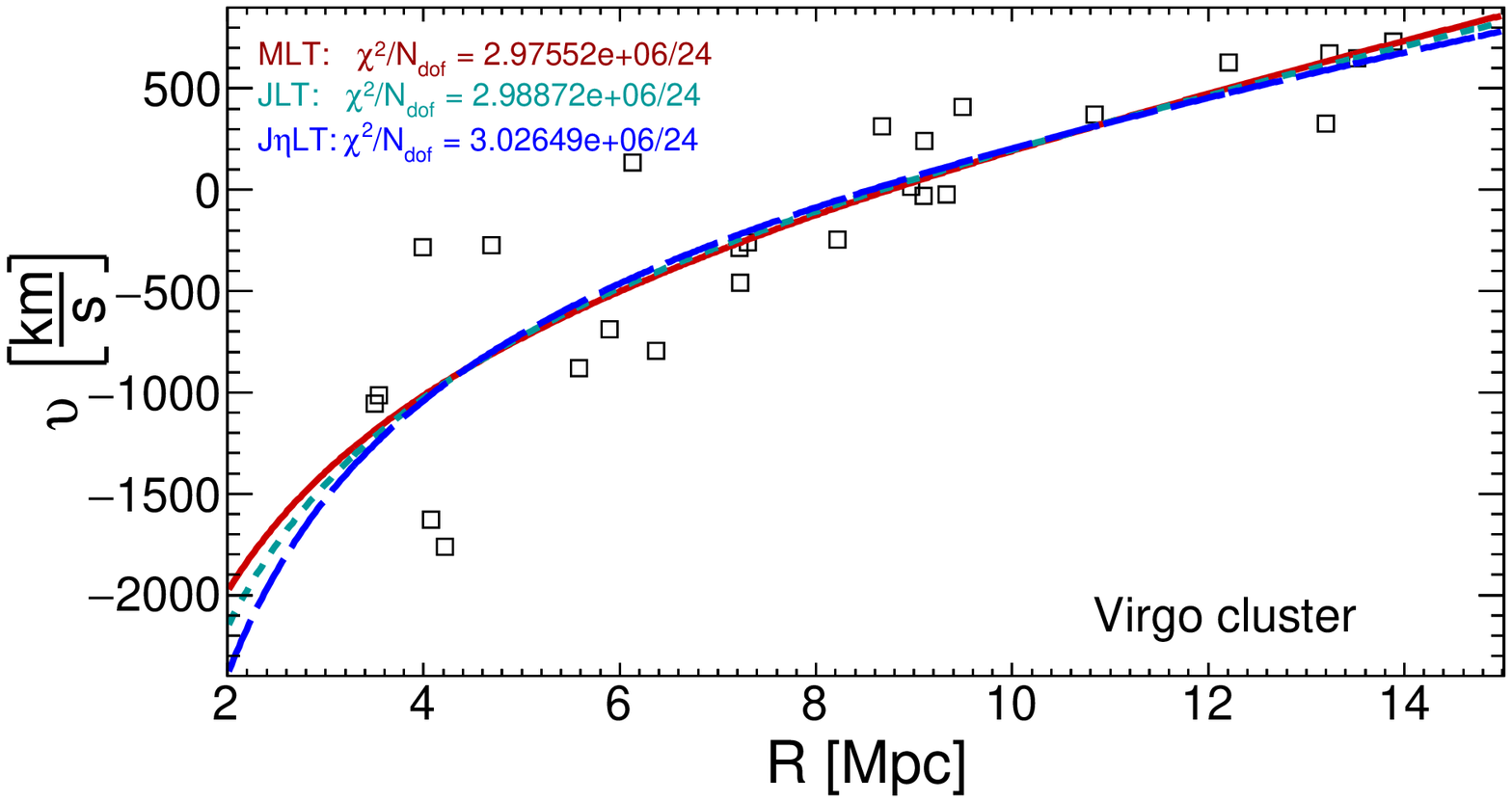}     
 \caption{
 Velocity-distance plots for the M31-MW group (top left panel), the M81 group (top right panel), the NGC 253 group (central left panel), the IC 342 group (central right panel), the CenA/M83 group (bottom left panel), the Virgo cluster (bottom right panel). The black dots are the data from Ref.\cite{Peirani2006,Peirani2008}.  Red solid lines correspond to the fit with the MLT model, \eqref{eq:Om}, cyan short dashed lines -- JLT model, \eqref{eq:j}, blue dashed lines -- J$\eta$LT model, \eqref{eq:eta}. The $\chi^{2}/N_{dof}$ values were added into each panel only in illustrative purpose, since the exact uncertainties of the data unknown.	
}
\label{fig:fig3}
\end{figure*}

Fig.\ref{fig:fig3} plots the $v-R$ relationships for the groups studied: the M31-MW group (top left panel), the M81 group (top right panel), the NGC 253 group (central left panel), the IC 342 group (central right panel), the CenA/M83 group (bottom left panel), the Virgo cluster (bottom right panel). The red squared are the data from Refs.\cite{Peirani2006,Peirani2008}.

%=================================================
\begin{table*}[ht]
  \centering
  \caption{
  Characteristic parameters of the examined groups. The rows 1-3 represent the value of the Hubble parameters for the MLT, JLT, and J$\eta$LT models. The rows 4-6 the masses of the groups in units of $10^{12} M_{\odot}$ for the same cases, and the rows -9 the values of the turn-around radius, $R_0$, in Mpc, for the same cases.  The last three rows give the velocity dispersion resulting from the fit of data to the $v-R$ relation for the same cases.
}
\begin{tabular}{lcccccc}
\scriptsize{}
&\scriptsize{\bf{M31/MW}}
&\scriptsize{\bf{M81}}
&\scriptsize{\bf{NGC 253}} 
&\scriptsize{\bf{IC 342}}
&\scriptsize{\bf{CenA/M83}}
&\scriptsize{\bf{Virgo}}\\
\midrule
\scriptsize{h($\Omega_{\Lambda}=0.7$)}
&\scriptsize{$0.73 \pm 0.04$}  &\scriptsize{$0.68 \pm 0.04$}  &\scriptsize{$0.63 \pm 0.06$} &\scriptsize{$0.58 \pm 0.10$}  &\scriptsize{$0.57 \pm 0.04$} &\scriptsize{$0.71 \pm 0.08$} \\
\scriptsize{h(j)} 
&\scriptsize{$0.70 \pm 0.04$}  &\scriptsize{$0.65 \pm 0.04$}  &\scriptsize{$0.63 \pm 0.06$}&\scriptsize{$0.56 \pm 0.10$}  &\scriptsize{$0.55 \pm 0.04$}  &\scriptsize{$0.65 \pm 0.09$} \\  
\scriptsize{h(j,$\eta$)} 
&\scriptsize{$0.69 \pm 0.04$}  &\scriptsize{$0.65 \pm 0.04$} &\scriptsize{$0.63 \pm 0.05$}  &\scriptsize{$0.55 \pm 0.10$}  &\scriptsize{$0.55 \pm 0.04$}&\scriptsize{$0.59 \pm 0.09$} \\ 
\hdashline[2.5pt/5pt]
\scriptsize{M($\Omega_{\Lambda}=0.7$) [$10^{12} M_{\odot}$]}
&\scriptsize{$2.49 \pm 0.50$}  &\scriptsize{$1.14 \pm 0.10$}  &\scriptsize{$0.14 \pm 0.15$}&\scriptsize{$0.22 \pm 0.12$} &\scriptsize{$2.16 \pm 0.50$} &\scriptsize{$1493 \pm 200$} \\
\scriptsize{M(j) [$10^{12} M_{\odot}$]}
&\scriptsize{$3.090 \pm 0.50$} &\scriptsize{$1.320 \pm 0.10$}  &\scriptsize{$0.195 \pm 0.10$} &\scriptsize{$0.263 \pm 0.10$}&\scriptsize{$2.655 \pm 0.50$}&\scriptsize{$1585 \pm 200$} \\ 
\scriptsize{M(j,$\eta$) [$10^{12} M_{\odot}$]} 
&\scriptsize{$3.570 \pm 0.40$}  &\scriptsize{$1.398 \pm 0.10$} &\scriptsize{$0.244 \pm 0.10$} &\scriptsize{$0.292 \pm 0.10$}&\scriptsize{$3.015 \pm 0.40$}&\scriptsize{$1525.55 \pm 200$} \\ 
\hdashline[2.5pt/5pt]
\scriptsize{$R_{0}(\Omega_{\Lambda}=0.7)$ [Mpc]}
&\scriptsize{$1.038 \pm 0.10$} &\scriptsize{$0.840 \pm 0.05$} &\scriptsize{$0.440 \pm 0.10$} &\scriptsize{$0.540 \pm 0.09$}&\scriptsize{$1.160 \pm 0.08$} &\scriptsize{$8.850 \pm 0.80$}\\
\scriptsize{$R_{0}(j)$ [Mpc]}
&\scriptsize{$1.04 \pm 0.10$}  &\scriptsize{$0.81 \pm 0.05$} &\scriptsize{$0.44 \pm 0.10$} &\scriptsize{$0.53 \pm 0.09$} &\scriptsize{$1.14 \pm 0.08$}&\scriptsize{$8.67 \pm 0.80$}\\
\scriptsize{$R_{0}(j,\eta)$ [Mpc]}
&\scriptsize{$1.02 \pm 0.10$} &\scriptsize{$0.78 \pm 0.05$} &\scriptsize{$0.44 \pm 0.10$} &\scriptsize{$0.52 \pm 0.09$}&\scriptsize{$1.13 \pm 0.08$} &\scriptsize{$8.56 \pm 0.80$}\\
\hdashline[2.5pt/5pt]
\scriptsize{$\sigma(\Omega_{\Lambda}=0.7)$ [km/s]}
&\scriptsize{$37.3$} &\scriptsize{$51.16$} &\scriptsize{$45.58$} &\scriptsize{$33.49$}&\scriptsize{$44.72$} &\scriptsize{$352.1$}\\
\scriptsize{$\sigma(j)$ [km/s]}
&\scriptsize{$38.3$}  &\scriptsize{$53.24$} &\scriptsize{$45.8$} &\scriptsize{$33.93$} &\scriptsize{$44.75$}&\scriptsize{$352.88$}\\
\scriptsize{$\sigma(j,\eta)$ [km/s]}
&\scriptsize{$38.8$} &\scriptsize{$54.77$} &\scriptsize{$45.9$} &\scriptsize{$34.48$}&\scriptsize{$44.81$} &\scriptsize{$355.1$}\\
\bottomrule   
    \end{tabular}%
\label{tab:hRM_fits}%
\end{table*}%
%=================================================

%%%%%%%%%%%%%%%%%%%%%%%%%%%%%%%%%%%%%%%%%%%%%%%%
%%%%%%%%%%%%%%%%%%%%%%%%%%%%%%%%%%%%%%%%%%%%%%%%
\subsection*{M31-MW}

We applied Eqs. (\ref{eq:Om}-\ref{eq:j}), and \eqref{eq:eta} to the Ref.\cite{Peirani2006} data. Both the mass and the Hubble parameter were allowed to vary. The results are shown in Table \ref{tab:hRM_fits}. Ref.\cite{Karachentsev2002} estimated a turn-around radius of $0.94 \pm 0.1$ Mpc and using the SLT model obtained a mass of $1.5 \times 10^{12} M_{\odot}$, which is much smaller than the estimate of Ref.\cite{Peirani2006} $ (2.5 \pm 0.7) \times 10^{12} M_{\odot}$, $R_0=1.0 \pm 0.1$ Mpc, and $h=0.74 \pm 0.04$. The value of the mass is larger than that of  Ref.\cite{Karachentsev2002}, that used the SLT model. A tendency of the LT models is that of giving higher masses, and smaller $h$ if the effect of the cosmological constant, angular momentum, and other effects which contribute with positive terms in the equation of motion are taken into account. In fact, Ref.\cite{Peirani2006} found a values of $h=0.87 \pm 0.05$ when using the SLT model, and $h=0.73 \pm 0.04$ in the MLT case.

The values of $R_0$, in all three cases (MLT, JLT, and J$\eta$LT) are in agreement, within the estimated uncertainties, with estimate reported in Ref.\cite{Peirani2006}.
Our values of $h$, and $M$ are in agreement to that of Ref.\cite{Peirani2006} in the MLT, JLT, cases, while in the J$\eta$LT the value is slightly larger. The average value of $h$ is smaller in the JLT, and J$\eta$LT models, while the reverse happens to the mass. We recall that the errors, come from the fitting procedure.

\subsection*{The M81 group}
The M81 group was studied by Refs.\cite{Karachentsev2002,Karachentsev2002a,Karachentsev2006}. The authors found $R_0 =0.89 \pm 0.05$ Mpc, smaller than our estimates, and $M=(1.03 \pm 0.17) \times 10^{12} M_{\odot}$, in agreement only with our MLT case. 
Ref.\cite{Peirani2008} found $M=(0.92 \pm 0.24) \times 10^{12} M_{\odot}$, smaller than our cases JLT, and J$\eta$LT
and $h=0.67 \pm 0.04$, in agreement with all our cases. 
Our JLT, and J$\eta$LT model estimates, as in the previous, and in all cases, gives average values of the mass, $M$ larger than the average of the estimates of Refs.\cite{Karachentsev2006,Peirani2008}.

\subsection*{The NGC253 group}   
Concerning this group, Ref.\cite{Karachentsev2003b} obtained $R_0=0.7 \pm 0.1$ Mpc, smaller than our estimates,
and $M=(5.5 \pm 2.2) \times 10^{11} M_{\odot}$, larger than our estimates. Ref.\cite{Peirani2008} found $M=(1.3 \pm 1.8) \times 10^{11} M_{\odot}$ whose larger uncertainties is probably due to incompleteness in the data. They also found $h=0.63 \pm 0.06$. Both their estimates for $h$, and $M$, are in agreement with all our cases. 

\subsection*{The IC342 group}   
According to Ref.\cite{Karachentsev2003a}, the group has $R_0 =0.9 \pm 0.1$, and $M=(1.07 \pm 0.33) \times 10^{12} M_{\odot}$, both larger than our estimates. Ref.\cite{Peirani2008} found a smaller value of the mass, $M=(2.0 \pm 1.3) \times 10^{11} M_{\odot}$, and also $R_0$ ($\simeq 0.53$ Mpc), while $h=0.57 \pm 0.10$. Our values of mass, turn-around radius, and $h$ agree with Ref.\cite{Peirani2008} estimates.      

\subsection*{The CenA/M83 group}
This group was studied by Refs.\cite{Karachentsev2002b,Karachentsev2007}. From the distances, and velocities of the group member, taking into account the cosmological constant they found $R_0=1.55 \pm 0.13$ Mpc, and $M=(6.4 \pm 1.8) \times 10^{12} M_{\odot}$, larger than our estimates. Ref.\cite{Woodley2006}, using different mass indicators found a larger mass ($M=(9.2 \pm 3) \times 10^{12} M_{\odot}$). Ref.\cite{Peirani2008}, found values 3-4 times smaller ($M=(2.1 \pm 0.5) \times 10^{12} M_{\odot}$), and $h=0.57 \pm 0.04$. In our analysis, both $M$, and $h$ are in agreement with Ref.\cite{Peirani2008}. 

\subsection*{The Virgo cluster}
Concerning Virgo, several estimates for the mass were done by means of the SLT model Ref.\cite{Hoffman1980,Fouque_2001}, by means of the Virial theorem Ref.\cite{Tully1984} finding masses smaller than $10^{15} M_{\odot}$, except Ref.\cite{Fouque_2001} who found a value of $1.3 \times 10^{15} M_{\odot}$. 
Using the MLT model Ref.\cite{Peirani2006} found $M=(1.10 \pm 0.12) \times 10^{15} M_{\odot}$, $h=0.65 \pm 0.09$, and $R_0= 8.6 \pm 0.8$ Mpc. Our estimates are in agreement with the value of $h$, $R_0$ of Ref.\cite{Peirani2006}, while the masses in the cases JLT, and J$\eta$LT are larger than in Ref.\cite{Peirani2006}.

In summary, our estimates usually agree with the estimates of Ref.\cite{Peirani2006,Peirani2008}, especially in the case MLT, and JLT. In some cases there are discrepancies with the predictions of our J$\eta$LT model. Moving from the SLT model to the MLT, JLT, and J$\eta$LT, the values of the cosmological constant decreases, and the opposite happens to the mass, $M$. 

Another important issue that is shown by Table \ref{tab:hRM_fits}, is that the values of $h$ are in some cases smaller than the known estimates. In the past decade or so, has been performed dozens of measurements of the Hubble constant, to try to overcome the Hubble constant tension.
As clear shown from Ref.\cite{Freedman2019}, from the year 2000 the constraints have changed from $72^{+8}_{-8}$ km/Mpc s, to the range $67-75$ km/Mpc s. Recent constraints from the gravitational wave signal of GW170817
gives $70.3^{+5.3}_{-5.0}$ km/Mpc s Ref.\cite{Hotokezaka2019}, $67.4^{+1.1}_{-1.2}$ km/Mpc s ($\rm DES+BAO+BBN$), and 
$67.5 \pm 1.1$ km/Mpc s Ref.\cite{Schoneberg2019}.
The previous constraints are in agreement with our results, except for CenA/M83 having $H=59$ km/Mpc s. The last discrepancy with observations may be due to non completeness of the data used in 2008 by Ref.\cite{Peirani2008}. 
Based on a large-scale survey of the Centaurus group done in 2014-2015, a significant amount of faint dwarf galaxy candidates were discovered Ref.\cite{Muller2017}. Therefore, the old data used in Ref.\cite{Peirani2008} may contain some selection bias so that the resulting $H$ obtained is systematically smaller.

%====================================================
%====================================================

%%%%%%%%%%%%%%%%%%%%%%%%%%%%%%%%%%%%%%%%%%%%%%%%
%%%%%%%%%%%%%%%%%%%%%%%%%%%%%%%%%%%%%%%%%%%%%%%%
\section*{Effects of cosmological constant, angular momentum, and dynamical friction}
\label{sec:EffectCosmologicalConst}

As we wrote in the Introduction, the mass predicted by the LT model is given by \eqref{eq:LT},
namely
\begin{equation} 
\label{eq:sandage}
M=\frac{\pi^2 R_0^3}{8GT_0^2}= 3.06\times 10^{12} h^2 R_0^3 M_{\odot}
\end{equation}
 
For the MLT, the value of $A$ can be obtained combining \eqref{eq:x}, and \eqref{eq:y}, and  one gets $A=3.6575$. By the definition of $A=\frac{2GM}{H_o^2 R_o^3}$, and recalling that in the $\Lambda$CDM model $H_0=f(\Omega)/T_0$, where 
\begin{equation}
f(\Omega)=\int_0^\infty \frac{dz}{(1+z)\sqrt{\Omega_{\Lambda}+\Omega_{\rm m} (1+z)^3}}
\end{equation}
we obtain, for $\Omega_{\Lambda}=0.7$
\begin{equation}
\label{eq:m1}
M=\frac{1.82875 H_0^2 R_0^3}{G}=  \frac{1.69945 R_0^3}{G T_0^2}=4.22\times 10^{12} h^2 R_0^3 M_{\odot}
\end{equation} 

Comparing \eqref{eq:LT} (or \eqref{eq:sandage}), and \eqref{eq:m1}, we get a difference of 38\%. 

In the case of the JLT model, with $K_j=0.78$ the value of $A$ is 5.037, then
\begin{equation} 
\label{eq:m2}
M=\frac{2.5185 H_0^2 R_0^3}{G}=\frac{2.3404 R_0^3}{G T_0^2}=5.8148 \times 10^{12} h^2 R_0^3 M_{\odot}
\end{equation}
and then the difference with the case LT is 90\%. Finally, in the case of the J$\eta$LT model, $A=6.05$
\begin{equation}
\label{eq:m3}
M=\frac{ 3.025 H_0^2 R_0^3}{G}= \frac{2.8111 R_0^3}{G T_0^2}=6.9843 \times 10^{12} h^2 R_0^3 M_{\odot}
\end{equation}
which means that the mass in this case is more than double of the case LT. The difference in mass between the previous cases is due to the modification of the perturbation evolution due to the effect of angular momentum, and dynamical friction as also shown in several papers Refs.\cite{DelPopolo1998,DelPopolo1999,DelPopolo2000,DelPopolo2006,DelPopolo2006a,DelPopolo2006b,DelPopolo2017}.

\begin{figure}[t]
	\centering 
	\includegraphics[scale=0.45]{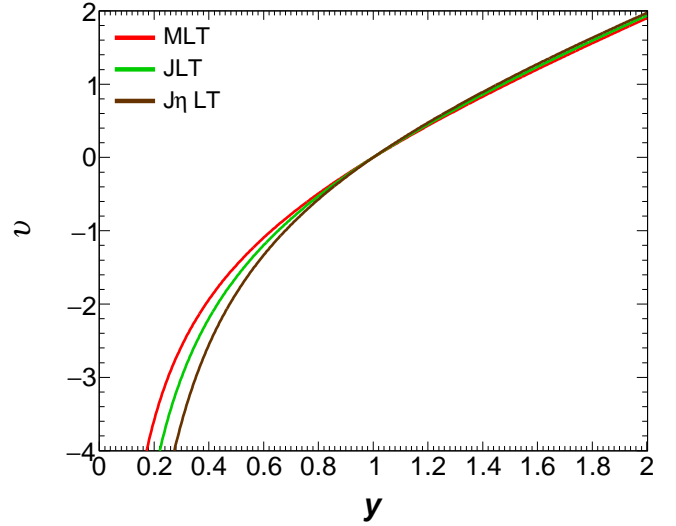}
	\caption{
		$v-R$ relationship for the MLT model (red curve), JLT (green curve), and J$\eta$LT (brown curve). 
	}
	\label{fig:fig4}
\end{figure}

The relation between mass, $M$, and turn-around radius, $R_0$, may be obtained also from \eqref{eq:Om}, \eqref{eq:j}, and \eqref{eq:eta}, solving the equation $v(R_0)=0$ with respect to $M$. In the case, LT, $A=2.655$, and the $v-R$ relation is given by 
\begin{equation}
\label{eq:lt00}
v(R)=-1.038 \frac{GM}{R}+1.196 H_0 R
\end{equation}
and 
\begin{equation}
\label{eq:lt0}
M=3.065 \times 10^{12} h^2 R_0^3 M_{\odot}
\end{equation}
In Fig.\ref{fig:fig4}, we plot the $v-y(R)$ relations for the MLT, the JLT, and the J$\eta$LT cases. 
For distances smaller than $R_0$, the plot shows that the J$\eta$LT cases gives larger negative velocities than the JLT model, and this larger negative velocities than the MLT model. This imply that the turn-around happens earlier in J$\eta$LT with respect to the JLT model, and similarly the turn-around happens earlier in JLT with respect to the MLT model. 
One interesting point is that the mass obtained from the $M-R_0$ relation in the case SLT (\eqref{eq:lt00}) is smaller than that
of the MLT case  (\eqref{eq:Om}). The last is smaller than the mass obtained with the JLT (\eqref{eq:j}), and this is smaller that that of J$\eta$LT case (\eqref{eq:eta}).

For example, fitting the data by means of \eqref{eq:lt00} (case SLT), the mass is $\simeq 10\%$ smaller than that obtained with \eqref{eq:Om} (case MLT)\footnote{The differences between the SLT, and MLT given by the $M-R_0$ relations (Eqs.(\ref{eq:sandage}), (\ref{eq:m1}))
is 38\%. }.  
Fitting the data by means of \eqref{eq:lt00}, and \eqref{eq:eta} the mass differences become larger (10\% in the case of M81, 100\% in the case of NGC 253, and around 40\% in the other cases, excluding Virgo).

The differences between the two methods can be explained as follows. In the method based on the fitting, the turn-around is obtained through $R_0=(\frac{2GM}{A H_0^2})^{1/3}$, and depends from $M$, and $H$, obtained through the fit.  

In the method based on the $M-R_0$ relation, $R_0$ is obtained by any method allowing the determination of this quantity, and then the $M-R_0$ relationship gives the mass. 

Another interesting point, is the decrease of $h$ from the SLT model, to the J$\eta$LT model. The maximum differences for the groups and clusters studied is  $\simeq 30\%$. 

%=================================================
%=================================================   

\begin{table}[!ht]
	\centering		
	\begin{tabular}{@{}ll}
		Stable structure  & range of $w$  \\
		\midrule
		M81     &$w \ge -1.5$\\
		IC342   &$w \ge -1$\\
		NGC253  &$w \ge -1$\\
		CenA/M83    &$w \ge -1.5$\\
		Local Group &$w \ge -2$\\
		Virgo       &$w \ge -1.5$\\
		\bottomrule
	\end{tabular}
	\caption{The allowed ranges of $w$.}
	\label{tab:table3}	
\end{table}

%%%%%%%%%%%%%%%%%%%%%%%%%%%%%%%%%%%%%%%%%%%%%%%%
%%%%%%%%%%%%%%%%%%%%%%%%%%%%%%%%%%%%%%%%%%%%%%%%
\section*{Constraints on the DM EoS parameter}
\label{sec:DMconstraints}
 
Recently, the turn-around radius, $R_0$ has been proposed as a promising way to test cosmological models Ref.\cite{Lopes2018}, DE, and disentangle between $\Lambda$CDM model, DE, and MG models Refs.\cite{Pavlidou2014,Pavlidou2014a,Faraoni2015,Bhattacharya2017,Lopes2018,Lopes2019}. 

Ref.\cite{Pavlidou2014} calculated $R_0$ for $\Lambda$CDM, and Ref.\cite{Pavlidou2014a} did the same for smooth DE. According to Ref.\cite{Lopes2019} $R_0$ is affected by modified gravity (MG) theories. In MG theories Ref.\cite{Capozziello2019} found a general relation for $R_0$, and 
Ref.\cite{Faraoni2015} found a method to get the same quantities in generic gravitational theories. In Ref.\cite{DelPopolo2020}, we used an extended spherical collapse model (ESCM) introduced, and adopted in Refs.\cite{DelPopolo2013,DelPopolo2013a,Pace2014,Mehrabi2017,Pace2019}, to show how $R_0$ is modified by the presence of vorticity, and shear in the equation of motion. We also showed how the $M-R_0$ plane can be used to put some constraints on the DE EoS parameter $w$, similarly to Refs.\cite{Pavlidou2014,Pavlidou2014a}. The constraints on $w$ depends on the estimated values of the mass and $R_0$ of galaxies, groups, and clusters. Some data where taken from Ref.\cite{Pavlidou2014a}, and others from Ref.\cite{Peirani2006,Peirani2008}.

With the revised value of mass, $M$, and $R_0$ presented in this paper, we recalculate the constraints showed in Ref.\cite{DelPopolo2020}. 

\begin{figure*}[!ht]
	\centering
	\includegraphics[scale=1.1]{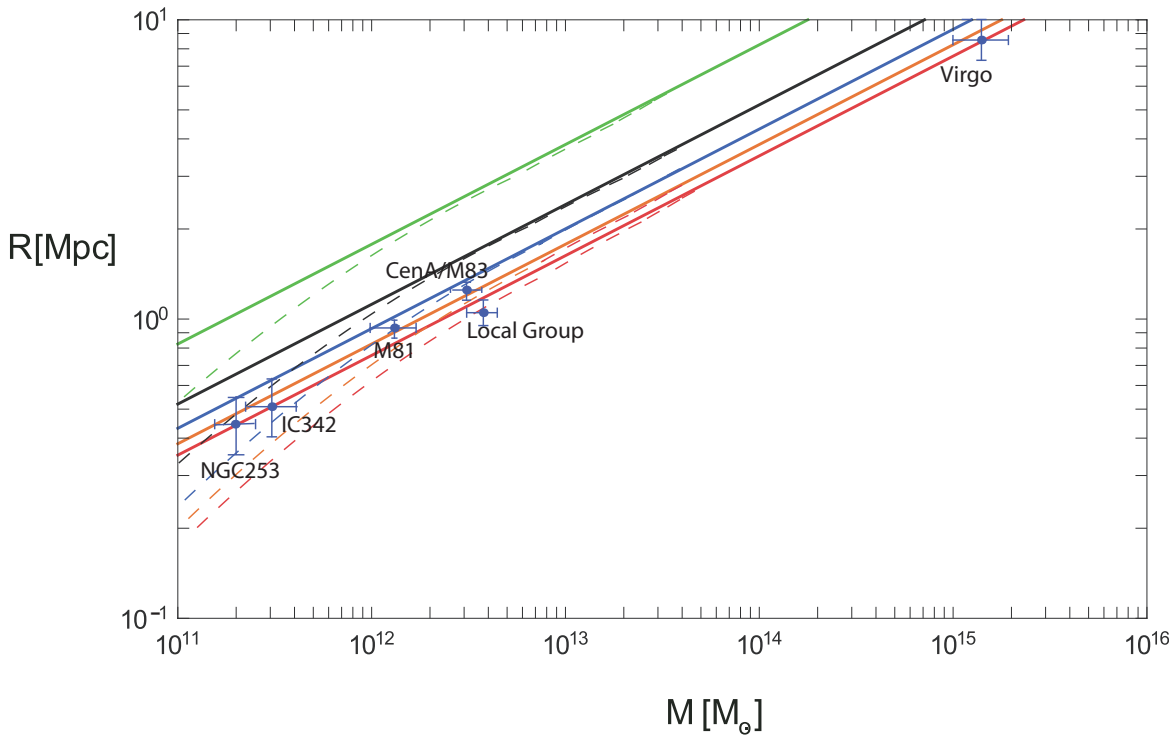}
	\caption{
Mass-radius relation of stable structures for different $w$. The solid lines from top to bottom represent $w=-0.5$ (solid green line), -1 (black solid line), -1.5 (blue solid line) ,-2 (pink solid line), -2.5 (red solid line). The dashed lines are the same of the previous lines, but they are obtained in Ref.\cite{DelPopolo2020}. The dots with error bars taken from Table~\ref{tab:hRM_fits}.
	}
	\label{fig:comparison}
\end{figure*}

Fig.\ref{fig:comparison} plots the mass-radius relation of stable structures for different $w$. The solid lines from top to bottom represent $w=-0.5$ (solid green line), -1 (black solid line), -1.5 (blue solid line) ,-2 (pink solid line), -2.5 (red solid line). The dashed lines are the same of the previous lines, but they are obtained using the model from Ref.\cite{DelPopolo2020}. The dots with error bars, are data obtained in the previous sections, and reported in Table~\ref{tab:hRM_fits} (case J$\eta$LT).

 The constraints to $w$ are reproduced in Table~\ref{tab:table3}. They are different from previous ones obtained by Ref.\cite{Pavlidou2014,Pavlidou2014a} based on the calculation of the mass, $M$, and $R_0$ by means of the virial theorem or the LT model. 

%=================================================   
%=================================================

%%%%%%%%%%%%%%%%%%%%%%%%%%%%%%%%%%%%%%%%%
%%%%%%%%%%%%%%%%%%%%%%%%%%%%%%%%%%%%%%%%%
\section*{Conclusions}
\label{sec:Conclusions}

In this paper, we extended the modified LT (MLT) model Refs.\cite{Peirani2006,Peirani2008} to take account the effect of angular momentum and dynamical friction. The inclusion of these two quantities in the equation of motion modifies the evolution of perturbations as described by the MLT model. The collapse of shells inside the zero-velocity surface collapse earlier when adding the angular momentum (JLT model), and dynamical friction term (J$\eta$LT model). After solving the equation of motion, we got the relationships between mass, $M$, and the turn-around radius $R_0$, similar to those obtained for the SLT model by Ref.\cite{Sandage1986}, and for the MLT model by Refs.\cite{Peirani2006,Peirani2008}. 
The relationships show, for a given $R_0$, a larger mass of the perturbation when angular momentum, and dynamical friction are taken into account. If one can obtain by some method the value of the turn-around, these relations show that the perturbation mass is 90\% (JLT model), and two times larger (J$\eta$LT model) with respect to the SLT model. 
In the paper, we also found velocity, $v$, radius, $R$, relationships for the cases considered depending on mass and the Hubble constant. These were fitted to the data of the local group, M81, NGC 253, IC342, CenA/M83, and Virgo. The values of the masses obtained fitting the data by means of \eqref{eq:eta} (J$\eta$LT model) are larger than those obtained by means of  \eqref{eq:lt00} (SLT model). The mass difference is 10\% in the case of M81, 100\% in the case of NGC 253, and around 40\% in the other cases.

The Hubble parameter becomes smaller when introducing angular momentum, and dynamical friction with respect to the SLT model. The same happens when adding the cosmological constant to the SLT model, as noticed by Refs.\cite{Peirani2006,Peirani2008}. 

Finally, we used the mass, $M$, and $R_0$ for the studied objects to put constraints to $w$. The constraints obtained differ from those obtained in previous papers Refs.\cite{Pavlidou2014,Pavlidou2014a} based on the calculation of the mass, $M$, and $R_0$ by means of the virial theorem or the LT model.

%\showmatmethods{} % Display the Materials and Methods section

\acknow{
The authors wish to express their gratitude to S. Peirani and A. De Freitas Pacheco for a fruitful discussion.
}

\showacknow{} % Display the acknowledgments section

% Bibliography
\bibliography{corrected_MasterBib}

\end{document}